\documentclass[10pt, conference, compsocconf]{IEEEtran}

\usepackage{times}
\usepackage{epsfig}
\usepackage{graphicx}
\usepackage{amsmath}
\usepackage{amssymb}
\usepackage{subfig}

\usepackage{authblk}
\ifCLASSINFOpdf
\else
\fi
\begin{document}
%
% paper title
% can use linebreaks \\ within to get better formatting as desired
\title{Concurrent Validity of Automatic Speech and Pause Measures During Passage Reading in ALS}

\author[1,2]{Saeid Alavi Naeini}
\author[1]{Leif Simmatis}
\author[1,3]{Yana Yunusova}
\author[1,2,4,5]{Babak Taati}
\affil[1]{Kite Research Institute, Toronto Rehabilitation Institute -- University Health Network}
\affil[2]{Institute of Biomedical Engineering, University of Toronto, Toronto, ON, Canada}
% 32608, United States}
\affil[3]{Department of Speech-Language Pathology, University of Toronto, Toronto, Canada}
\affil[4]{Department of Computer Science, University of Toronto, Toronto, ON, Canada}
\affil[5]{Vector Institute, Toronto, ON, Canada}
% \affil[1]{\authorcr Email: {\tt saeid.alavi@mail.utoronto.ca}}
% \affil[2]{\authorcr Email: {\tt Leif.Simmatis@uhn.ca}}
% {
% % \makeatletter
% % \renewcommand\AB@affilsepx{ \protect\Affilfont}
% % \makeatother

% % \affil[ ]{}

% \makeatletter
% \renewcommand\AB@affilsepx{, \protect\Affilfont}
% \makeatother

% \affil[ ]{\tt saeid.alavi@mail.utoronto.ca, \tt Leif.Simmatis@uhn.ca,}
% }
% {
% \makeatletter
% \renewcommand\AB@affilsepx{ \protect\Affilfont}
% \makeatother

% \affil[ ]{}

% \makeatletter
% \renewcommand\AB@affilsepx{, \protect\Affilfont}
% \makeatother
% \affil[ ]{\tt deniz.jafari@mail.utoronto.ca, \tt dguarinlopez@fit.edu,}
% }
% {
% \makeatletter
% \renewcommand\AB@affilsepx{ \protect\Affilfont}
% \makeatother

% \affil[ ]{}

% \makeatletter
% \renewcommand\AB@affilsepx{, \protect\Affilfont}
% \makeatother

% \affil[ ]{\tt yana.yunusova@utoronto.ca, \tt Babak.Taati@uhn.ca}
% }

% use for special paper notices
%\IEEEspecialpapernotice{(Invited Paper)}

% make the title area
\maketitle

\begin{abstract}
The analysis of speech measures in individuals with amyotrophic lateral sclerosis (ALS) can provide essential information for early diagnosis and tracking disease progression. However, current methods for extracting speech and pause features are manual or semi-automatic, which makes them time–consuming and labour–intensive. The advent of speech-text alignment algorithms provides an opportunity for inexpensive, automated, and accurate analysis of speech measures in individuals with ALS. There is a need to validate speech and pause features calculated by these algorithms against current gold standard methods. In this study, we extracted 8 speech/pause features from 646 audio files of individuals with ALS and healthy controls performing passage reading. Two pretrained forced alignment models -- one using transformers and another using a Gaussian mixture / hidden Markov architecture  -- were used for automatic feature extraction. The results were then validated against semi-automatic speech/pause analysis software, with further subgroup analyses based on audio quality and disease severity. Features extracted using  transformer-based forced alignment had the highest agreement with gold standards, including in terms of audio quality and disease severity. This study lays the groundwork for future intelligent diagnostic support systems for clinicians, and for novel methods of tracking disease progression remotely from home.
\end{abstract}


% For peer review papers, you can put extra information on the cover
% page as needed:
% \ifCLASSOPTIONpeerreview
% \begin{center} \bfseries EDICS Category: 3-BBND \end{center}
% \fi
%
% For peerreview papers, this IEEEtran command inserts a page break and
% creates the second title. It will be ignored for other modes.
\IEEEpeerreviewmaketitle

\makeatletter
\def\thickhline{%
  \noalign{\ifnum0=`}\fi\hrule \@height \thickarrayrulewidth \futurelet
   \reserved@a\@xthickhline}
\def\@xthickhline{\ifx\reserved@a\thickhline
               \vskip\doublerulesep
               \vskip-\thickarrayrulewidth
             \fi
      \ifnum0=`{\fi}}
\makeatother

\newlength{\thickarrayrulewidth}
\setlength{\thickarrayrulewidth}{4\arrayrulewidth}

%%%%%%%%% BODY TEXT
\section{Introduction}
\begin{table*}[h]
\renewcommand{\arraystretch}{1.45}
\centering
\caption{Spearman correlation ($\rho$) between features calculated from the Wav2Vec2 method and from the SPA software, divided based on audio quality. The number of audio files in each category is presented in parenthesis}
\begin{tabular}{l c rrrrrr}  % creating 8 columns
\thickhline
 &\multicolumn{2}{c}{\textbf{Good (314)}} & \multicolumn{2}{c}{\textbf{Fair (120)}} &\multicolumn{2}{c}{\textbf{Poor (212)}} %&\multicolumn{2}{c}{\textbf{Rest (200)}}
\\
\cline{2-7}
&\multicolumn{1}{c}{$\rho$} & \multicolumn{1}{c}{p} &\multicolumn{1}{c}{$\rho$} & \multicolumn{1}{c}{p}&\multicolumn{1}{c}{$\rho$} & \multicolumn{1}{c}{p}
% &\multicolumn{1}{c}{$\rho$} & \multicolumn{1}{c}{p}
\\
\thickhline             % inserts single-line% Entering 1st row 
\textbf{\textit{Pause duration}} &$0.82$ & $<.001$ & $0.84$ & $<.001$ & $0.62$ & $<.001$ 
% & $0.71$ & $<.001$
\\
\textbf{\textit{Total duration}} &$0.98$ & $<.001$ & $0.96$ & $<.001$ & $0.93$ & $<.001$
% & $0.90$ & $<.001$
\\
\textbf{\textit{Speech duration}} &$0.96$ & $<.001$ & $0.91$ & $<.001$ & $0.70$ & $<.001$ 
% & $0.82$ & $<.001$
\\
\textbf{\textit{Pause event}} &$0.73$ & $<.001$ & $0.80$ & $<.001$ & $0.58$ & $<.001$
% & $0.70$ & $<.001$
\\
\textbf{\textit{$\%$ Pause}} &$0.72$ & $<.001$ & $0.75$ & $<.001$ & $0.48$ & $<.001$
% & $0.57$ & $<.001$
\\
\textbf{\textit{Mean phrase}} &$0.77$ & $<.001$ & $0.70$ & $<.001$ & $0.50$ & $<.001$
% & $0.59$ & $<.001$
\\
\textbf{\textit{CV phrase duration}} &$-0.05$ & $0.39$ & $0.12$ & $0.18$ & $0.01$ & $0.86$ %& $0.21$ & $.02$
\\
\textbf{\textit{CV pause duration}} &$0.53$ & $<.001$ & $0.38$ & $<.001$ & $0.27$ & $<.001$ %& $0.21$ & $.02$
\\
\thickhline                         % inserts single-line
\end{tabular}
\label{table1}
\end{table*}

\begin{table*}[h]
\renewcommand{\arraystretch}{1.45}
\centering
\caption{Spearman correlation ($\rho$) between features calculated from the MFA model and from the SPA software, divided based on audio quality. The number of audio files in each category is presented in parenthesis}
\begin{tabular}{l c rrrrrr}  % creating 8 columns
\thickhline
  &\multicolumn{2}{c}{\textbf{Good (314)}} & \multicolumn{2}{c}{\textbf{Fair (120)}} &\multicolumn{2}{c}{\textbf{Poor (212)}}
  %&\multicolumn{2}{c}{\textbf{Rest (200)}}
\\
\cline{2-7}
&\multicolumn{1}{c}{$\rho$} & \multicolumn{1}{c}{p} &\multicolumn{1}{c}{$\rho$} & \multicolumn{1}{c}{p}&\multicolumn{1}{c}{$\rho$} & \multicolumn{1}{c}{p}

% &\multicolumn{1}{c}{$\rho$} & \multicolumn{1}{c}{p}
\\ 
\thickhline             % inserts single-line% Entering 1st row 
\textbf{\textit{Pause duration}} &$0.77$ & $<.001$ & $0.74$ & $<.001$ & $0.52$ & $<.001$
% & $0.68$ & $<.001$
\\
\textbf{\textit{Total duration}} &$0.90$ & $<.001$ & $0.86$ & $<.001$ & $0.85$ & $<.001$ 
% & $0.88$ & $<.001$
\\
\textbf{\textit{Speech duration}} &$0.87$ & $<.001$ & $0.86$ & $<.001$ & $0.76$ & $<.001$ 
% & $0.81$ & $<.001$
\\
\textbf{\textit{Pause event}} &$0.74$ & $<.001$ & $0.64$ & $<.001$ & $0.56$ & $<.001$ 
% & $0.65$ & $<.001$
\\
\textbf{\textit{$\%$ Pause}} &$0.65$ & $<.001$ & $0.62$ & $<.001$ & $0.40$ & $.04$ %& $0.62$ & $<.001$
\\
\textbf{\textit{Mean phrase}} &$0.60$ & $<.001$ & $0.50$ & $<.001$ & $0.58$ & $<.001$ 
%& $0.47$ & $<.001$
\\
\textbf{\textit{CV phrase duration}} &$-0.05$ & $.51$ & $-0.38$ & $.05$ & $0.02$ & $.86$ %&$0.19$ & $<.001$
\\
\textbf{\textit{CV pause duration}} &$0.32$ & $<.001$ & $0.31$ & $.11$ & $0.27$ & $.11$ %&$0.26$ & $<.001$
\\
\thickhline                         % inserts single-line
\end{tabular}
\label{table2}
\end{table*}
Amyotrophic lateral sclerosis (ALS) is a rapidly progressing neurodegenerative disorder that destroys motor neurons in the cerebral cortex, brainstem, and spinal cord~\cite{kiernan2011amyotrophic}. The emergence of bulbar signs is an essential milestone in the progression of ALS, as it causes the loss of speech intelligibility and, consequently, has a significant impact on the quality of life of people with ALS. The loss of communication, especially speech, has been reported as one of the worst aspects of the disease~\cite{hecht2002subjective}.

There are established clinical assessment techniques for diagnosing bulbar symptoms associated with ALS and tracking disease severity; however, these have important limitations. Clinical impressions of range of motion, speed, and symmetry of oral musculature are valuable, but these assessments are typically subjective~\cite{shellikeri2016speech}. The reliability of these clinician-based judgements have been questioned and lack of reliable assessment tools contributes to long diagnostic delay in ALS (up to 18 months) ~\cite{turner2010diagnostic}. Earlier detection of bulbar symptoms in ALS could facilitate finding augmentative and alternative communication referrals, and assist early bulbar symptom management~\cite{green2013bulbar}. There is a substantial need to develop accessible, automated systems that can objectively and reliably assess bulbar ALS symptoms in nonclinical settings.

% Speech and pause features, often measured using the Speech Intelligibility Test (SIT)~\cite{dorsey2007speech}, have been recommended as preferred objective measures

Speech timing can be broadly measured using the Speech Intelligibility Test (SIT) which provides a measure of speaking rate~\cite{dorsey2007speech}. Speaking rate has been recommended as the preferred objective measure of bulbar decline by speech-language pathologists~\cite{yorkston1993speech, ball2001protocol}. Passage reading tasks are excellent alternatives to the SIT, since they can also provide measures of speech and pause duration~\cite{green2004algorithmic}. Studies have shown that these features can distinguish bulbar disease at different stages, with or without respiratory deficits~\cite{yunusova2016profiling}. 

Current methods of obtaining speech and pause features are either manual or semi-automatic, which require time-intensive processing and training of professional staff~\cite{barnett2020reliability}. Forced alignment techniques -- which temporally align passage text and speech audio signals -- could potentially overcome these limitations. Past attempts to automatically parse and align dysarthric speech have been error-prone~\cite{pleva2015automatic, yeung2015improving}. 

Transformers are a new class of deep learning models that have achieved state-of-the-art performance in a variety of tasks, e.g. in natural language processing or computer vision~\cite{vaswani2017attention,dosovitskiy2020image}. The recent development of robust and accurate transformer-based forced alignment models~\cite{baevski2020wav2vec} presents a new opportunity for fully automated speech/pause feature extraction in dysarthric speech. However, it is an open question as to whether novel methods could replace current gold standards for dysarthric speech alignment.

The aim of this study is to establish the concurrent validity of automatic speech-text alignment algorithms against the Speech and Pause Analysis (SPA) software, which is a validated semi-automated pause identification tool~\cite{barnett2020reliability}. We investigate two open-source forced alignment models -- including a new transformer-based model~\cite{baevski2020wav2vec} -- and report correlations between 8 speech and pause features calculated via SPA vs. via forced alignment. Our hypothesis is that features calculated via automatic text-speech alignment will have  strong relationship 
% (Spearman $\rho>.5$)  
with their SPA counterparts. 

% page 3 tables
\begin{table*}[h]
\renewcommand{\arraystretch}{1.5}
\centering
\caption{Spearman correlation ($\rho$) between features calculated from the Wav2Vec2 model on Good audio quality data and from the SPA software, divided based on disease severity. The number of audio files in each category is presented in parenthesis}
\begin{tabular}{l c rrrrrrr}  % creating 8 columns
\thickhline
 &\multicolumn{2}{c}{\textbf{HC (147)}} & \multicolumn{2}{c}{\textbf{Mild (70)}} &\multicolumn{2}{c}{\textbf{Moderate (59)}} &\multicolumn{2}{c}{\textbf{Severe (38)}}
\\ [0.5ex]
\cline{2-9}
&\multicolumn{1}{c}{$\rho$} & \multicolumn{1}{c}{p} &\multicolumn{1}{c}{$\rho$} & \multicolumn{1}{c}{p}&\multicolumn{1}{c}{$\rho$} & \multicolumn{1}{c}{p}&\multicolumn{1}{c}{$\rho$} & \multicolumn{1}{c}{p}
\\ [0.5ex] 
\thickhline             % inserts single-line% Entering 1st row 
\textbf{\textit{Pause duration}} &$0.82$ & $<.001$ & $0.84$ & $<.001$ & $0.66$ & $<.001$ & $0.94$ & $<.001$\\
\textbf{\textit{Total duration}} &$0.97$ & $<.001$ & $0.95$ & $<.001$ & $0.96$ & $<.001$ & $0.99$ & $<.001$\\
\textbf{\textit{Speech duration}} &$0.93$ & $<.001$ & $0.93$ & $<.001$ & $0.90$ & $<.001$ & $0.98$ & $<.001$\\
\textbf{\textit{Pause event}} &$0.53$ & $<.001$ & $0.73$ & $<.001$ & $0.60$ & $<.001$ & $0.92$ & $<.001$\\
\textbf{\textit{$\%$ Pause}} &$0.74$ & $<.001$ & $0.82$ & $<.001$ & $0.54$ & $<.001$ & $0.83$ & $.001$\\
\textbf{\textit{Mean phrase}} &$0.76$ & $<.001$ & $0.80$ & $<.001$ & $0.68$ & $<.001$ & $0.91$ & $<.001$\\
\textbf{\textit{CV phrase duration}} &$0.04$ & $.61$ & $-0.25$ & $.03$ & $0.04$ & $.60$ & $-0.10$ & $.53$\\
\textbf{\textit{CV pause duration}} &$0.32$ & $<.001$ & $0.59$ & $<.001$ & $0.67$ & $<.001$ & $0.74$ & $<.001$\\[1ex]
\thickhline                         % inserts single-line
\end{tabular}
\label{table3}
\end{table*}

\begin{table*}[h]
\renewcommand{\arraystretch}{1.5}
\centering
\caption{Spearman correlation ($\rho$)  between features calculated from the MFA model on Good audio quality data and from the SPA software, divided based on disease severity. The number of audio files in each category is presented in parenthesis}
\begin{tabular}{l c rrrrrrr}  % creating 8 columns
\thickhline
 &\multicolumn{2}{c}{\textbf{HC (147)}} & \multicolumn{2}{c}{\textbf{Mild (70)}} &\multicolumn{2}{c}{\textbf{Moderate (59)}} &\multicolumn{2}{c}{\textbf{Severe (38)}}
\\ [0.5ex]
\cline{2-9}
&\multicolumn{1}{c}{$\rho$} & \multicolumn{1}{c}{p} &\multicolumn{1}{c}{$\rho$} & \multicolumn{1}{c}{p}&\multicolumn{1}{c}{$\rho$} & \multicolumn{1}{c}{p}&\multicolumn{1}{c}{$\rho$} & \multicolumn{1}{c}{p}
\\ [0.5ex] 
\thickhline             % inserts single-line% Entering 1st row 
\textbf{\textit{Pause duration}} &$0.64$ & $<.001$ & $0.78$ & $<.001$ & $0.51$ & $<.001$ & $0.75$ & $<.001$\\
\textbf{\textit{Total duration}} &$0.86$ & $<.001$ & $0.92$ & $<.001$ & $0.85$ & $<.001$ & $0.44$ & $.07$\\
\textbf{\textit{Speech duration}} &$0.92$ & $<.001$ & $0.66$ & $.002$ & $0.65$ & $.002$ & $0.63$ & $.007$\\
\textbf{\textit{Pause event}} &$0.52$ & $<.001$ & $0.74$ & $<.001$ & $0.77$ & $<.001$ & $0.70$ & $.002$\\
\textbf{\textit{$\%$ Pause}} &$0.68$ & $<.001$ & $0.77$ & $<.001$ & $0.41$ & $.07$ & $0.41$ & $.09$\\
\textbf{\textit{Mean phrase}} &$0.52$ & $<.001$ & $0.74$ & $<.001$ & $0.70$ & $<.001$ & $0.39$ & $.12$\\
\textbf{\textit{CV phrase duration}} &$-0.02$ & $.83$ & $0.13$ & $.60$ & $-0.08$ & $.73$ & $-0.10$ & $.71$\\
\textbf{\textit{CV pause duration}} &$0.15$ & $.14$ & $0.56$ & $.01$ & $0.53$ & $.02$ & $0.43$ & $.08$\\[1ex]
\thickhline                         % inserts single-line
\end{tabular}
\label{table4}
\end{table*}

\begin{figure*}[t]
\centering

\includegraphics[width=1\linewidth]{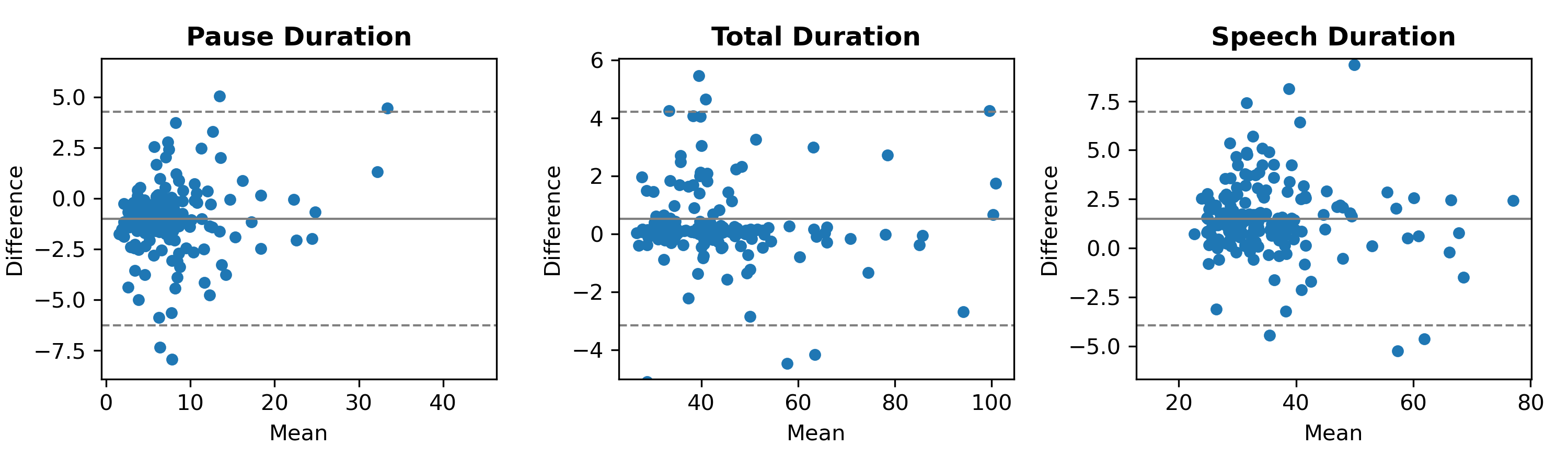}%
 
\caption{The Bland–Altman plots between Wav2Vec2 and SPA features of \textit{pause duration}, \textit{speech duration}, and \textit{total duration} on Good audio quality data. The dashed lines are the lower and upper limits of agreement (1.96$\times$standard deviation) and the solid line is zero.}
\label{fig_blant_altman}%
\end{figure*}

\section{Methods}
\subsection{Participants}

We used a dataset of 646 speech recordings, obtained from 3 observational longitudinal studies of individuals with ALS and healthy controls (HC). The recordings were of the participants reading a short (60 word) passage -- the bamboo passage~\cite{green2004algorithmic,yunusova2016profiling}. 

The total number of participants was 462 (ALS: 243, controls: 219). Additional demographic and clinical details are presented in~\cite{barnett2020reliability}. The time interval between recordings varied from 3 to 6 months when more than one recording was available from a participant. Individuals with ALS were diagnosed with possible, probable, or definite ALS, as defined by the El Escorial Criteria~\cite{brooks2000escorial}.

Speaking rate was obtained for each speaker and session using the SIT and was calculated as the number of words per minute (WPM). Similar to~\cite{shellikeri2016speech}, we classified the individuals with ALS into 3 groups of mild, moderate, and severe based on their speaking rate. The mild group comprised of individuals with speaking rates greater than 160 WPM; the moderate group included individuals with speaking rates between 120 and 160 WPM; and the severe group had a speaking rate less than 120 WPM. 

We also categorized the quality of audio files into 3 groups of Good, Fair, and Poor based on signal-to-noise ratio (SNR) of each acoustic file. Following previous research~\cite{xu2004simple}, audio files with SNR below 15 dB were labeled as Poor, 15--20 dB were labeled as Fair, and above 20 dB were labeled as Good. The reason for having different quality groups were due to recordings being obtained in a clinical (in contrast to a laboratory) context, resulting in environmental noise and quiet/muffed recordings in some cases.  
% Recordings significantly corrupted by background noise or chatter were excluded. 

% The final dataset consisted of 446 passage recordings from individuals with ALS.
% All the non-annotated audio files were added to a fourth group called 'Rest'.

\subsection{Procedure}

We applied 2 different forced alignment techniques for the automatic parsing and speech/text alignment of the Bamboo
passage - a standard passage used for the analysis of bulbar disease severity in ALS~\cite{green2004algorithmic}. All recordings were also previously parsed using the semi-automated SPA software.
% The first forced alignment method calculates the similarity score between two audio sequences using Dynamic Time Warping (DTW) and uses Mel-frequency cepstral coefficients (MFCCs) as
% acoustic feature. we used a python library called Aeneas~\cite{pettarin2020practical} to perform the DTW-based speech-text alignment. In preliminary analysis we found that AWS Polly Text-to-Speech (TTS) API worked best as the TTS engine wrapper. 
The first forced alignment method was the Montreal Forced Aligner (MFA)~\cite{mcauliffe2017montreal}. The automatic speech recognition pipeline in MFA is based on a Gaussian mixture/hidden Markov (GMM/HMM) architecture from the open-source Kaldi speech recognition toolkit~\cite{mcauliffe2017montreal}. MFA uses Mel-frequency cepstral coefficients (MFCCs) as
acoustic input features. 
For the second forced alignment method, we used a pretrained transformer-based Wav2Vec2 model~\cite{baevski2020wav2vec} from the Charsiu library~\cite{zhu2021phone}. The model can perform both text-dependent and text-independent phone-to-audio alignment and maintains good performance in different settings. In all cases, we performed speech-text alignment in a supervised manner, i.e. by providing the text of bamboo passage as an input to the model.

\subsection{Feature Extraction}

The forced alignment output comprised the time stamps associated with each speech and pause event greater than 300 milliseconds. We then extracted the same 8 features used in~\cite{barnett2020reliability} based on their ability to distinguish individuals with ALS from controls. They included:
\begin{enumerate}
\item \textit{Pause duration} (sec): total duration of all pause events (excluding the start and end pauses).
\item \textit{Total duration} (sec):  recording time from the onset of the first sentence to the end of the last sentence.
\item \textit{Speech duration} (sec): total duration of
all speech events (pauses excluded).
\item \textit{Pause events}: number of pauses while reading the passage.
\item \textit{$\%$ Pause}: percentage of total reading time spent pausing (only pauses between speech events). 
\item \textit{Mean phrase} (sec): average
duration of a phrase. Phrases were defined as sections of continuous speech without a pause.
\item \textit{CV phrase duration}: Coefficient of variation of phrase durations (a normalized measure of variability).
\item \textit{CV pause duration}: Coefficient of variation of pause durations.
\end{enumerate}

\subsection{Evaluation}

Spearman’s correlation coefficients ($\rho$) were used to determine the agreement between the features extracted from forced alignment algorithms and from the SPA software. We compared these sets of 8 features across 3 different conditions: 1) two forced alignment algorithms (MFA, and Wav2Vec2), 2) four severity levels (HC, and ALS mild, moderate, or severe), and 3) three audio qualities (Good, Fair, Poor). The  correlation coefficients ($\rho$) and p-values are reported (significance threshold was p$<$0.05). 
% In this analysis, we considered Spearman correlation ($\rho$) values in the $.40-.59$ range as moderate, in the $.60-.79$ range as strong, and in the $.80-1.0$ range as very strong association.
%While we were not able to measure and compare the accuracy of the Wav2Vec2 algorithm, due to not having the manual parsing of audio files, 
We also present the  agreement between a subset of the Wav2Vec2 features and their SPA counterparts using Bland–Altman plots. The subset of 3 features was selected as those that were previously reported to have the highest area under the receiver operating characteristic curve (AUC-ROC) in detecting bulbar changes in otherwise bulbar presymptomatic
individuals with ALS
~\cite{barnett2020reliability}.

\section{Results}

The results of correlation analysis for the transformer-based (Wav2Vec2) and HMM-based (MFA) models are outlined in Tables~\ref{table1}~and~\ref{table2}, respectively. Performance is separately reported for  recordings based on their audio quality. 
Tables~\ref{table3} and ~\ref{table4} present the correlation results for Wav2Vec2 and MFA models based on disease severity. This analysis is performed on Good quality data to ensure disease severity is the only variable affecting the correlation. 

The Bland–Altman plots for \textit{pause duration}, \textit{speech duration}, and \textit{total duration} are shown in Figure~\ref{fig_blant_altman}. The plots illustrate the agreement between features extracted using the Wav2Vec2 model vs. using the SPA software.
For all three features, there is similar agreement between Wav2Vec2 and SPA features as all 3 plots have relatively narrow limits of agreement band ($\pm$1.96$\times$standard deviation) which contains 95$\%$ of the values. 
% In total duration and speech duration cases, the points on the Bland–Altman plots were uniformly and tightly scattered around the horizontal axis. In pause duration case, it seems that there is an increase in variability of the differences as the magnitude of the measurement increases. 
% The majority of the points were inside the agreement band for the total duration plot. This is inline with correlation results as total duration had the highest agreement amongst all features.

\section{Discussion}

The current gold standard for the parsing of clinical audio data and feature extraction is the manual or semi-automatic approach, which is time–consuming and labour–intensive. Here, we compared two novel, automated methods of audio-based parsing to SPA, which has been previously validated for parsing clinical audio data~\cite{green2004algorithmic}. We demonstrated that, although both MFA and Wav2Vec2 performed well with respect to SPA, Wav2Vec2 generalized better across clinical severities. However, both methods were sensitive to audio data quality. The present work represents an important step toward developing automated tools for assessing neurodegenerative disorders, including ALS, in the home setting.

The general pattern than can be observed in the correlation results is that the Wav2Vec2 model performed better with most features being very strongly (or strongly) associated. We also observed that correlations between SPA and automated methods appeared to change as a function of audio sample quality (Tables~\ref{table1} and~\ref{table2}), with Good audio having the strongest correlations. This agrees with the results of previous studies; for example, the performance of forced alignment algorithms is reduced when sound to noise ratio is not optimized~\cite{mcauliffe2017montreal, zhu2021phone}. Our results reinforce the need for high quality audio data in the assessment of neurological and neurodegenerative diseases. 
% Since the Rest category contains a mixture of audio qualities, we can see that its correlation values are somewhere between the values in Good and Poor quality categories. 

% also add that: (very very brief, one or two sentences max
% 1. many features are very strongly (or strongly) associated
% 2. pattern: Wav2Vec2 genrally better
% (can copy some text from the commented paragraph starting with $BBBBBBBB$

The present results add to existing literature and support previous calls to incorporate more flexible methods into automated and semi-automated speech analysis. In ~\cite{tanchip2022validating}, automated diadochokinetic analysis methods were explored, and it was found that automated methods' performance suffered as clinical severity increased. The results of the present study suggest that this could be alleviated by using flexible and robust methods, such as Wav2Vec2-based forced alignment in our current passage-reading paradigm. Importantly, we observed that correlations between SPA/Wave2Vec2 across clinical severities (Tables 3 and 4) were almost uniformly better than they were in the SPA/MFA comparison. This suggests that transformer-based models are more robust to range of clinical severities and are viable replacements for current manual and semi-automatic procedures used for clinical audio data analysis.

In conclusion, our work has demonstrated that automated forced alignment methods perform well compared to a semi-automated ground truth, and that the Wav2Vec2-based approach in particular is robust across a range of clinical severity levels. Importantly, our results also highlight the need for good-quality audio data to be used for clinical speech analyses. These results have important implications for the development of new clinical tools for home-based speech assessment, and also for future work that will aim to prognose patients based on speech function. These are challenging problems ~\cite{kimura2006progression, lee1995prognosis} that will require further future work. It will also be important in future work to extend our current findings to include other populations that experience speech impairments, such as stroke and Parkinson’s disease.

\section{Acknowledgements}

The authors would like to thank the Natural Sciences and Engineering Research Council of Canada (NSERC), ALS Canada, Parkinson Canada, Canadian Partnership for Stroke Recovery, Michael J. Fox Foundation and Weston Brain Institute, the Kite Research Institute -- UHN, and Mitacs for supporting our research. We would also like to thank Dr. Madhura Kulkarni for her role in supporting data collection and management for this project.

% trigger a \newpage just before the given reference
% number - used to balance the columns on the last page
% adjust value as needed - may need to be readjusted if
% the document is modified later
%\IEEEtriggeratref{8}
% The "triggered" command can be changed if desired:
%\IEEEtriggercmd{\enlargethispage{-5in}}

% references section

% can use a bibliography generated by BibTeX as a .bbl file
% BibTeX documentation can be easily obtained at:
% http://www.ctan.org/tex-archive/biblio/bibtex/contrib/doc/
% The IEEEtran BibTeX style support page is at:
% http://www.michaelshell.org/tex/ieeetran/bibtex/
%\bibliographystyle{IEEEtran}
% argument is your BibTeX string definitions and bibliography database(s)
%\bibliography{IEEEabrv,../bib/paper}
%
% <OR> manually copy in the resultant .bbl file
% set second argument of \begin to the number of references
% (used to reserve space for the reference number labels box)
{\small
\bibliographystyle{IEEEtranBST/IEEEtran}
\bibliography{IEEEtranBST/IEEEfull}
}

% that's all folks
\end{document}